\documentclass{article}
\usepackage{amsmath,amssymb,cite}

\def\beq{\begin{eqnarray}}
\def\eeq{\end{eqnarray}}
\def\bsp{\begin{split}}
\def\esp{\end{split}}
\def\dx{{\bf dx}}
\def\dt{{\bf dt}}
\def\dy{{\bf dy}}
\def\dz{{\bf dz}}
\def\dw{{\bf dw}}
\def\isom{\mathrm{Isom}}

\newcommand{\mf}[1]{{\mathfrak #1}}

\newcommand{\mb}[1]{{\mathbb #1}}
\newcommand{\mc}[1]{{\mathcal #1}}
\newcommand{\mbold}[1]{\mbox{\boldmath${#1}$}}

\begin{document}

\title{\textbf{Einstein metrics: Homogeneous solvmanifolds,
    generalised Heisenberg groups and Black Holes}}
\author{\textbf{Sigbj\o rn Hervik}\thanks{S.Hervik@damtp.cam.ac.uk} \\
DAMTP, \\
Centre for Mathematical Sciences,\\
Cambridge University\\
Wilberforce Rd. \\
Cambridge CB3 0WA, UK}
\maketitle
\begin{abstract}
In this paper we construct Einstein spaces with negative Ricci 
curvature in various dimensions. These spaces -- which can be thought of as generalised AdS
spacetimes --  can be classified in terms of the geometry of
the horospheres in Poincar\'e-like coordinates, and can be both homogeneous
and static. By using simple building blocks, which in general are
homogeneous Einstein solvmanifolds, we give a general 
algorithm for constructing Einstein metrics where the horospheres are any
product of generalised Heisenberg geometries, nilgeometries,
solvegeometries, or 
Ricci-flat manifolds. Furthermore, we show that
all of these spaces can give rise to black
holes with the horizon geometry corresponding to the geometry of the
horospheres, by explicitly deriving their metrics.  
\end{abstract}

\section{Introduction} 
In the recent years the study of Anti-de Sitter spaces (AdS)
 has been intense. Because they arise as maximally symmetric
solutions to the Einstein equations with a negative cosmological
constant they were thought for a long time to be irrelevant to physics
 and merely a 
mathematical curiosity. However, from a mathematical point of view,
negatively curved spaces have an incredibly rich structure
\cite{thurston,thur:97,bp}. For example, one of the biggest problems in
  classifying three-manifolds is the enormous number of compact
 hyperbolic spaces; in general, one finds the negatively curved spaces
 have an enormously
 diverse variety. For the AdS spaces -- which can be thought of as
 the Lorentzian versions of the hyperbolic spaces -- this diversity
 can, for example, be seen
 in the many  AdS black holes that one knows
of \cite{btz,abhp,hp,blp,Mann}. 

AdS spaces have also come in the focus of research after the advent of
superstring theory \cite{BF}. One of the most promising ideas is the
AdS/CFT correspondence \cite{Maldacena} which relates a supergravity theory
in the interior of AdS to a field theory in the boundary of AdS
space. This correspondence illustrates the interplay between the
structure of the interior of these spaces with the structure on the
conformal boundary of the AdS space. 

In this paper, we shall construct negatively curved Einstein manifolds
which are, unlike the AdS spaces, not maximally symmetric. However,
like the AdS spaces, a large class of them are homogeneous and
static, though in general they need not be. However, there will always
exist a non-trivial group acting on the space. This group manifests
itself in horospherical coordinates where the space is foliated into
horospheres. The Euclidean AdS
spaces (or real hyperbolic spaces if one likes) in horospherical
coordinates\footnote{These are sometimes also called Poincar\'e
coordinates.} are simply 
\beq
ds^2={\bf dw}^2+e^{-2w}\sum_{i=1}^n(\dx^i)^2.
\label{metricHyperbolic}\eeq
Here, the horospheres are given by $w=$ constant and are 
flat Euclidean spaces. In this paper, we shall construct spaces for which the
geometries of the horospheres are products of generalised Heisenberg
groups, nilgeometries or solvegeometries. They are negatively curved Einstein spaces
of the form  
\beq 
ds^2={\bf dw}^2+\sum_{i=1}^ne^{-2q_iw}({\mbold\omega^i})^2.
\label{eq:genmetric}\eeq
These spaces are
analogous to the AdS spaces and their Lorentzian versions exist in any
dimension  higher than four. Higher-dimensional gravity has  already given us  some
surprises; for example the great variety of black hole spacetimes (see
\cite{GH} for a review) and even black strings \cite{ER}. In this paper we shall add even more black
hole solutions to the myriad of known black hole spacetimes by finding
black holes with horizons modelled on an arbitrary product of
generalised Heisenberg groups, nilgeometries and
solvegeometries. These  black 
holes are not asymptotically AdS; they are
asymptotically of the form (\ref{eq:genmetric}). 

This paper is organised as follows. First we consider in detail 
complex hyperbolic spaces. These spaces are the simplest non-maximally
symmetric,  negatively curved Einstein spaces. In the analysis,
we discuss in detail how the horospheres can be equipped with a  Heisenberg
geometry.  Then, in 
section 3, we provide  the simplest non-trivial example and  solve
the field equations for a space where 
the horospheres are a product between a Ricci-flat manifold and a
Heisenberg geometry. The solution found is Einstein, and we discuss
its isometries and give
conditions for when the space is homogeneous. By considering a hypersurface in a product of Einstein
spaces, we show in section 4 how we can iteratively construct
higher-dimensional Einstein spaces by using simple building
blocks. These building blocks can be any manifold of a certain form,
and in particular, they can be any homogeneous Einstein solvmanifold. 
Examples of such homogeneous Einstein solvmanifolds are given in
section 5. Among these spaces are the so-called Damek-Ricci spaces which
have  horospheres equipped with generalised Heisenberg
geometries. Also, examples where the horospheres are nilgeometries and
solvegeometries are given. Lastly, we show that all spaces constructed by the
iterative procedure have  black hole analogues. We give explicit
metrics for those black hole solutions which can have horizon geometries
modelled on any product of generalised Heisenberg groups,
nilgeometries and solvegeometries, and a
Ricci-flat space. 

\section{Complex hyperbolic Spaces}
The metrics we are going to construct have many common features
with the complex hyperbolic spaces, $\mb{H}_{\mb{C}}^{n+1}$. In fact, the
construction is motivated from the existence and properties of these
spaces. We will first review some of the aspects of the complex
hyperbolic spaces  which heavily motivated our construction. In this
respect, the book by Goldman \cite{goldman} is an 
indispensable source and reference. 

The metric can be written in real horospherical coordinates
\beq
ds^2&=&\dw^2 +e^{-2w}\left[\dx+\frac
12\sum_{k=1}^n(y^k\dz^k-z^k\dy^k)\right]^2 \nonumber \\
&&+ e^{-w}\sum_{k=1}^n\left[(\dy^k)^2+(\dz^k)^2\right].
\label{metricHCN}\eeq
The metric is a K{\"a}hler metric with constant holomorphic curvature,
 and thus, is also Einstein. 
 This form of the metric is particularly useful for our
purposes as we shall see. The full isometry group of this metric is
$PU(n+1,1)$:
\[
PU(n+1,1)\equiv U(n+1,1)/\sim, \quad \text{where } {\sf A}\sim {\sf B} \Leftrightarrow
{\sf A}= \lambda{\sf B}, ~\lambda\in \mb{C}. \]
 At the
Lie algebra level this isometry group has the following Iwasawa decomposition:
\beq
\mf{g}=\bigoplus_{k=-2}^2\mf{g}_{k},
\eeq 
and 
\beq
\mf{n}_{\pm}\equiv\mf{g}_{\pm 1}\oplus\mf{g}_{\pm 2}
\eeq
defines two copies of the $(2n+1)$-dimensional Heisenberg algebra. More precisely, 
\beq
\mf{g}_0\cong \mf{u}(n)\times \mb{R}, \quad \mf{g}_{\pm 1}\cong \mb{C}^{n},\quad \mf{g}_{\pm 2}\cong \mb{R}, 
\eeq
and $\left[\mf{g}_{i},\mf{g}_{j}\right]=\mf{g}_{i+j}$. In
horospherical coordinates this decomposition is easy to interpret; the
Heisenberg algebra $\mf{n}_{-}$ generates a Heisenberg
group\footnote{We will use a notation where lower case Gothic
  letters correspond to the Lie algebra; upper case Gothic to the Lie
  group; and calligraphic letters to the corresponding geometry
  (i.e. the Lie group equipped with a left-invariant metric).}, $\mf{H}$,
which acts simply transitively on the horospheres $w=$constant. Hence, each horosphere has a natural associated
Heisenberg geometry, $\mc{H}$. Furthermore, $\mf{g}_0$ can be
given a geometric interpretation in terms of this Heisenberg geometry as well. $\mf{g}_0$ generates  $U(n)$-rotations,
and an $\mb{R}_+$-dilation with respect to the origin of $\mc{H}$. More
explicitly, introducing the complex column vector
\beq
{\mbold\zeta}=\begin{bmatrix}
y^1+iz^1,& y^2+iz^2, & \cdots & y^n+iz^n
\end{bmatrix}^T,
\eeq
the group $U(n)$ acts by matrix multiplication 
\beq
{\mbold\zeta}\overset{{\sf A}}{\longmapsto}{\sf A}{\mbold\zeta}.
\eeq
The isometry group of $\mathcal{H}$ is now given by the semi-direct
product
\beq
\isom(\mc{H})= U(n)\ltimes \mf{H}.
\eeq 
On the Heisenberg space with coordinates $(x,{\mbold\zeta})$ the dilation,
$\phi_{\lambda}$, 
acts as
\[ (x,{\mbold\zeta})\overset{\phi_{\lambda}}{\longmapsto} (e^{2\lambda}x,e^{\lambda}{\mbold\zeta}). \] 
which translates into an isometry of the metric (\ref{metricHCN}) by
acting along the $w$-coordinate: 
\[ (w,x,{\mbold\zeta})\overset{\phi_{\lambda}}{\longmapsto} (w+\lambda,
e^{2\lambda}x,e^{\lambda}{\mbold\zeta}). \]

The \emph{similarity group} of $\mc{H}$ is therefore the transformations
generated by $\mf{g}_0$ and the isometries $\mf{H}$:
\beq
\mathrm{Sim}(\mc{H})=\left(\mb{R}_+\times U(n)\right)\ltimes \mf{H}.
\label{SimH}\eeq

Note that the curve  defined by $(x,{\mbold\zeta})=$constant, is a geodesic from
$w=-\infty$ to $w=\infty$. The geodesic connects two point on the
boundary of $\mb{H}_{\mb{C}}^{n+1}$, denoted $\partial \mb{H}_{\mb{C}}^{n+1}$. The point at infinity, given by
$w=\infty$, corresponds to  one point on the boundary. We will
call this point $p_{\infty}$. The dilation maps the geodesic through
the origin of the Heisenberg geometry onto itself. The
algebra, $\mf{g}_0$, generates all the isometries leaving the geodesic
through the origin invariant. 

Interestingly, the isometries induce a similarity group  on the
boundary of the complex hyperbolic space in the following sense. The
boundary of $\mb{H}_{\mb{C}}^{n+1}$
is topologically a sphere $S^{2n+1}$. However, the isometries
generated by $\mf{g}_0$ and $\mf{H}$, induce -- via  relation
(\ref{SimH}) -- similarity transformations on the boundary minus the 
point at infinity: $\partial\mb{H}_{\mb{C}}^{n+1}-\{p_{\infty}\}$. The boundary
$\partial\mb{H}_{\mb{C}}^{n+1}-\{p_{\infty}\}$ can therefore be
considered as a Heisenberg geometry on which $\mf{g}_0$ and $\mf{H}$
acts as similarity transformations. In fact, the remaining isometries
of $\mb{H}_{\mb{C}}^{n+1}$
will act on the boundary as conformal transformations. However, in our
construction later on the remaining isometries will be broken; we will
squash the complex hyperbolic space slightly along the
$w$-coordinate. 

This feature of complex hyperbolic spaces  is very similar to the
real hyperbolic space.  In the real case, the horospheres have a
Euclidean geometry, $\mb{E}^{n}$,  and the similarity group 
$\mathrm{Sim}(\mc{H})$ is interchanged with the similarity group
$\mathrm{Sim}(\mb{E}^{n})$. This is also one of the many reasons why the
AdS spaces are so interesting: they are the Lorentzian
versions of real hyperbolic spaces. It is therefore interesting to
know whether it is possible to extend the interesting geometric
properties of the complex hyperbolic spaces to the Lorentzian case. 

\section{Some Einstein metrics of dimension $(2+2n+m)$}
By including  extra  dimensions it is possible to construct
spaces for which the structure (\ref{SimH}) remains intact. The
starting point is a squashed complex hyperbolic space but where we
keep the some of the properties which we discussed earlier. We
construct the Einstein metrics as follows: Assume
that $\mc{M}$ is an $m$-dimensional Ricci-flat manifold with metric $\widetilde{ds}_m^2$;
i.e.
\beq
\widetilde{ds}_m^2=\tilde{g}_{AB}{\bf d}\chi^A{\bf d}\chi^B, \qquad \widetilde{R}_{AB}=0.
\label{dstilde}\eeq
Then we  start with the metric ansatz where the above metric is
warped in the following way\footnote{Later, we shall give a different
  interpretation of this metric; namely as a hypersurface in a higher-dimensional space.}
\beq
ds^2&=&e^{-2pw}\widetilde{ds}_m^2+Q^2\dw^2 +e^{-2w}\left[\dx+\frac
12\sum_{k=1}^n(y^k\dz^k-z^k\dy^k)\right]^2 \nonumber \\
&&+ e^{-w}\sum_{k=1}^n\left[(\dy^k)^2+(\dz^k)^2\right] 
\label{metricHn}\eeq
where $Q$ and $p$ are constants. 

The metric ansatz immediately leads to a diagonal Ricci tensor (see
appendix) and the 
large number of symmetries makes the field equations particularly easy to
solve. For the values 
\beq
p=\frac{n+2}{2(n+1)},\qquad Q^2=1+\frac{m(n+2)}{2(n+1)^2}
\eeq
the metric (\ref{metricHn}) is an Einstein metric with 
\beq
R_{\mu\nu}=-\frac{n+2}{2}g_{\mu\nu}.
\eeq
Of course, by rescaling the metric (\ref{metricHn}), we can find a
metric with $R_{\mu\nu}=-\alpha^2 g_{\mu\nu}$
for any $\alpha^2 >0$.  Thus, in the
following we will consider 
one particular $\alpha^2$, but bear in mind that any rescaling is
possible. 

These spaces have many similarities with the AdS spaces and  some of their properties are directly related to the the complex hyperbolic space in horospherical
coordinates. The isometry group of  (\ref{metricHn}) depends on the
properties of $\mc{M}$, and in general the metric (\ref{metricHn})
will not be homogeneous. Note that if there exists a proper
similarity transformation of $\widetilde{ds}_m^2$, 
\beq
\psi_{\kappa}:~ \chi^A\mapsto \psi_{\kappa}(\chi^A) \nonumber \\
\psi_{\kappa}^*\widetilde{ds}_m^2=e^{2\kappa}\widetilde{ds}_m^2,
\eeq
where $\kappa$ is a non-zero constant, then the dilation, $\phi_{\lambda}$, can be extended to an isometry of
(\ref{metricHn}):
\beq
(\chi^A,x,{\mbold\zeta})\overset{\phi_{\lambda}}{\longmapsto} (\psi_{p\lambda}(\chi^A),e^{2\lambda}x,e^{\lambda}{\mbold\zeta}).
\eeq
This leads to the following observation: \emph{If ${\mc M}$ is a
  homogeneous space and, in addition, allows for a proper
  homothety, then the metric given by eq. (\ref{metricHn}) is
  homogeneous.} 

For the metric (\ref{metricHn}) to be homogeneous, the existence of a
proper homothety of ${\mc M}$ is crucial; however, there are
still quite a few spaces having these properties. For example, the
following Ricci flat spaces are homogeneous and possess a proper
homothety:
\begin{enumerate}
\item{} The Euclidean spaces $\mb{E}^m$, or the Minkowski spaces
  $\mb{M}^m$.
\item{} The $m$-dimensional Milne universes. 
\item{} The $m$-dimensional homogeneous vacuum plane-waves \cite{Blau}
  \footnote{See also \cite{Siklos} which has some interesting
  negatively curved Einstein
  spaces conformally equivalent to the vacuum
  plane-waves.}.  
\end{enumerate}
There are also examples of homogeneous spacetimes $\mathcal{M}$ which do not admit a
proper homothety, and inhomogeneous spacetimes admitting a
proper homothety.\footnote{See also \cite{GH} where examples
of compact Ricci-flat manifolds are provided. }

\section{Extending the scope: ``Prime decomposition''} 
We will here describe an iterative procedure of constructing
generalisations of the metric (\ref{metricHn}). It is based on a
simple observation regarding the Gauss' equation for hypersurfaces. 

Assume we have two negatively curved Einstein manifolds, $\mc{M}$, and
$\mc{N}$, with metrics of the form
\beq
ds^2_{\mc{M}}&=&{\bf dw}^2+\sum_{i=1}^me^{-2p_iw}\left({\mbold\omega^i}\right)^2,
\qquad R_{AB}=-\alpha^2g_{AB} \label{metricM} \\
ds^2_{\mc{N}}&=&{\bf
  dv}^2+\sum_{i=1}^ne^{-2q_iv}\left({\mbold\chi^i}\right)^2,\qquad
R_{ab}=-\beta^2 g_{ab}. \label{metricN}
\eeq
It is essential here that the one-forms ${\mbold\omega^i}$ and
${\mbold\chi^i}$ both form a closed algebra; i.e. all the $w$ and
$v$ dependence is in the exponential prefactor. Explicitly, this
means that the forms ${\mbold\omega}^i$ obey
\beq
{\bf d}{\mbold\omega}^i=-{\mbold\Omega}^i_{~j}\wedge {\mbold\omega}^j,
\eeq
and ${\mbold\Omega}^i_{~j}$ does not involve $w$. 

The construction now goes as follows. We define the product space
$\mc{M}\times\mc{N}$ and consider a  hypersurface
$\Sigma\subset\mc{M}\times\mc{N}$. Our aim is to tune the parameters
and choose the hypersurface such that the induced metric, $h_{\mu\nu}$, on $\Sigma$
is an Einstein metric. 

From the the theory of hypersurfaces, we have the well known result
(the contracted Gauss' equation, see e.g. \cite{wald})
\beq
{}^{(\Sigma)}R_{\mu\nu}={}^{(\mc{M}\times\mc{N})}R_{\alpha\beta}h^{\alpha}_{~\mu}h^{\beta}_{~\nu}-{}^{(\mc{M}\times\mc{N})}R^{\lambda}_{~\alpha\sigma\beta}n_{\lambda}n^{\sigma}h^{\alpha}_{~\mu}h^{\beta}_{~\nu}+KK_{\mu\nu}-K^{\alpha}_{~\mu}K_{\alpha\nu}.
\label{Gauss}\eeq
Here, $n^{\mu}$ and $K_{\mu\nu}$ are the normal vector field and the
extrinsic curvature to the hypersurfaces, respectively. 

Now using the metrics (\ref{metricM}) and (\ref{metricN}), the Ricci
tensor in $\mc{M}\times\mc{N}$ becomes block diagonal. Let us consider
the following hypersurface, 
\beq
v=\gamma w,
\eeq
where $\gamma$ is a constant. 
The unit normal is now given by 
\[ {\bf n}=-\frac{\gamma}{\sqrt{1+\gamma^2}}\frac{\partial}{\partial
  w}+\frac{1}{\sqrt{1+\gamma^2}}\frac{\partial}{\partial v}. \]

By calculating the Riemann tensor for $\mc{M}\times\mc{N}$ we get the
following for the metrics given by eqs. (\ref{metricM}) and
(\ref{metricN})
\[ {}^{(\mc{M}\times\mc{N})}R^{\lambda}_{~\alpha\sigma\beta}n_{\lambda}n^{\sigma}h^{\alpha}_{~\mu}h^{\beta}_{~\nu}=-K^{\alpha}_{~\mu}K_{\alpha\nu}.\]
Hence, Gauss' equation simplifies to 
\beq
{}^{(\Sigma)}R_{\mu\nu}={}^{(\mc{M}\times\mc{N})}R_{\alpha\beta}h^{\alpha}_{~\mu}h^{\beta}_{~\nu}+KK_{\mu\nu}.
\eeq
Note there are two simple choices for making $\Sigma$ an Einstein
space. In both cases we choose $\alpha^2=\beta^2$ (by rescaling the
metric). The first case
arises when $K_{\mu\nu}\propto h_{\mu\nu}$. However, as can be seen,
this happens only when $p_i=q_j\equiv p$. This is a well known example
and does not give any new solutions. The other
case is more interesting and is defined by the choice $K=0$. So, the following
will lead to an Einstein space:
\beq
\alpha^2=\beta^2, \quad K=0.
\eeq

Explicitly, the requirement $K=0$ leads to 
\beq
\gamma\sum_{i=1}^mp_i-\sum_{i=1}^nq_i =0,
\eeq
From this, $\gamma$ can be found and the metric on $\Sigma$ becomes
(after rescaling $w$)
\beq
ds^2_{\Sigma}&=& {\bf
  dw}^2+\sum_{i=1}^me^{-2P_iw}\left({\mbold\omega^i}\right)^2
+\sum_{i=1}^ne^{-2Q_iw}\left({\mbold\chi^i}\right)^2
\label{metricSigma}, \nonumber \\
&& P_i=\frac{1}{\sqrt{1+\gamma^2}}p_i, \quad  Q_i=\frac{\gamma}{\sqrt{1+\gamma^2}}q_i.
\eeq

Using this procedure, we can iteratively construct negatively curved
Einstein spaces in higher dimensions using building blocks of the form
(\ref{metricM}). Thus we have provided a product rule for the
geometry of the horosphere. Assume that $M(\mc{H}_1)$ and
$M(\mc{H}_2)$ are Einstein metrics of the above form with horospheres
equipped with the geometries $\mc{H}_1$ and $\mc{H}_2$,
respectively. Then we have a product rule 
\beq
M(\mc{H}_1)\odot M(\mc{H}_2)= M(\mc{H}_1\times \mc{H}_2),
\eeq
given by  $ M(\mc{H}_1)\odot M(\mc{H}_2)\cong \Sigma \subset M(\mc{H}_1)\times M(\mc{H}_2)$ with the induced
metric. Note that 
\[ \left[M(\mc{H}_1)\odot M(\mc{H}_2)\right]\odot
M(\mc{H}_3)=M(\mc{H}_1)\odot \left[M(\mc{H}_2)\odot
M(\mc{H}_3)\right]; \]
i.e. the operation $\odot$ is associative. In this way the
classification reduces to considering irreducible ``prime'' manifolds
$M(\mc{H}_i)$.

\subsection*{General formula}
In fact, we can also give a general formula for the exponents. Assume
that we have $N$ Einstein spaces, 
\beq
ds^2_{A}&=&{\bf dw}^2+\sum_{i=1}^{m_A}e^{-2p_{i(A)}w}\left({\mbold\omega_{(A)}^i}\right)^2,
\quad R_{\mu\nu}=-\alpha^2g_{\mu\nu}, \quad A=1..N.
\eeq
Then there is an Einstein space 
  \beq
ds^2&=&{\bf dw}^2+\sum_{A=1}^N\sum_{i=1}^{m_A}e^{-2q_{i(A)}w}\left({\mbold\omega_{(A)}^i}\right)^2,
\qquad R_{\mu\nu}=-\alpha^2g_{\mu\nu}
\eeq
where
\beq
q_{i(A)}=p_{i(A)}\left(\sum_{j=1}^{m_A}p_{j(A)}\right)\left[{\sum_{B=1}^n\left(\sum_{j=1}^{m_B}p_{j(B)}\right)^2}\right]^{-\frac
  12}.
\eeq

\section{Einstein solvmanifolds}

Note that we have assumed in the above construction that the metric
can be more general than the usual complex and real hyperbolic
spaces. Thus we may wonder if there are other ``building blocks'' than
the ones already considered  of the form
(\ref{metricM}). Indeed there are and, in fact, they are so numerous
that not all such  manifolds are 
known. However, here we will consider some generalisations of these
building blocks; 
namely homogeneous Einstein solvmanifolds.

Let us consider the homogeneous spaces for which the Lie algebras,
$\mf{s}$, obey (see e.g. \cite{Wolter}):
\begin{enumerate} 
\item{} The Iwasawa decomposition has the following orthogonal decomposition: 
\[ {\mf s}=\mf{a}\oplus\mf{n},\quad [\mf{s},\mf{s}]=\mf{n},\] 
where $\mf{a}$ is abelian, and $\mf{n}$ is nilpotent.
\item{} All operators $\mathrm{ad}_{X}$, $X\in\mf{a}$ are symmetric.
\item{} For some $X^0\in\mf{a}$,
  $\left.\mathrm{ad}_{X^0}\right|_{\mf{n}}$ has positive eigenvalues.
\end{enumerate}
These Lie algebras
give rise to group manifolds, and since ${\mf{s}}$ is a solvable
  algebra, these are so-called
  \emph{solvmanifolds}. By 
using the left-invariant one-forms we can turn the groups into
Riemannian homogeneous spaces. These spaces can always be put
onto the form (\ref{metricM}) and they are strong candidates for
Einstein spaces with negative curvature \cite{Wolter,Wolter2}. 

The different classes of these solvmanifolds are characterised by the
dimension of $\mf{a}$ and the property of the nilpotent part
$\mf{n}$. Some general results are known, but a complete list of
manifolds of this type is lacking. Here, due to the multiplication
rule as given above, we will only consider the cases where $\mf{s}$
leads to an irreducible solvmanifold $M(\mc{H})$.

\subsection{$\dim(\mf{a})=1$, and $\mf{n}$ abelian.}
These are the well-known hyperbolic spaces. The metric can be written
on the form given in eq. (\ref{metricHyperbolic}).

\subsection{$\dim(\mf{a})=1$, and $\mf{n}$ generalised Heisenberg algebra:
  Damek-Ricci spaces} 
In this case, we can construct all possible cases. The
nilpotent part $\mf{n}$ are generalised Heisenberg algebras,
and the extended solvable group give rise to  so-called Damek-Ricci spaces. 
\subsubsection*{Generalised Heisenberg algebras} 
The (ordinary) $(2n+1)$-dimensional Heisenberg algebra
is the nilpotent part of  the Iwasawa decomposition of
$\isom(\mb{H}_{\mb{C}}^{n+1})$. Here we will consider the 
\emph{generalised Heisenberg spaces}\footnote{See
  e.g. \cite{GHG}.}, $\mc{H}_{m,n}$. The generalised Heisenberg
algebras are defined as follows. 
Let $\mf{b}$ and $\mf{z}$ be real vector spaces of dimension $m$ and
$n$, respectively, such that $\mf{n}$ is
the orthogonal sum $\mf{n}=\mf{b}\oplus\mf{z}$. Assume that the
commutator is a map $[-,-]:\mf{n}\times\mf{n}\mapsto\mf{z}$ and
$[\mf{z},\mf{z}]=0$. Furthermore, assume that there exists a
$J:\mf{z}\mapsto{\rm End}(\mf{b}),~ Z\mapsto J_{Z}$ such that 
\[ \langle [X,Y],Z\rangle =\langle Y,J_ZX\rangle, \quad X,Y\in
\mf{b},~Z\in \mf{z}. \]
The triple $(\mf{b}, \mf{z}, J)$ defines uniquely a 2-step nilpotent
algebra $\mf{n}$ and, in fact, a 2-step nilpotent simply connected Lie
group with a left-invariant metric. We say that $\mf{n}$ is a
\emph{generalised Heisenberg algebra} if, in addition,
\[ J_Z^2=-\langle Z,Z\rangle {\rm Id}_{\mf{b}}, \quad \forall Z\in \mf{z}. \]

The generalised Heisenberg spaces,
$\mc{H}_{m,n}$, are the corresponding group manifolds 
equipped with  orthonormal left-invariant one-forms of the form  
\beq
{\mbold\omega}^A &=&\dx^A+\frac 12B^A_{ab}\ y^a{\bf dy}^b, \qquad
A=1...m \nonumber \\
{\mbold\omega}^a &=&\dy^a, \qquad a=1...n.
\eeq
Here are $B^A_{ab}$ antisymmetric in the lower indices. It should be
noted that, given an $m$, not all $n$'s are allowed. For example,
for $m=1$, which gives the ordinary Heisenberg spaces, only even $n$
are allowed. In general, the number $n$ can have the following values
\beq
n=k n_0, \quad k\in \mb{N}, 
\eeq
where $n_0$ is given in Table \ref{table}.
\begin{table}
\begin{tabular}{|c|c|c|c|c|c|c|c|c|}\hline
$m$ & $8p$ & $8p+1$& $8p+2$& $8p+3$ & $8p+4$ & $8p+5$& $8p+6$ & $8p+7$\\ \hline
$n_0$ & $2^{4p}$ & $2^{4p+1}$ &  $2^{4p+2}$ & $2^{4p+2}$ &  $2^{4p+3}$ &
$2^{4p+3}$ &  $2^{4p+3}$ & $2^{4p+3}$ \\ \hline
\end{tabular}
\caption{Generalised Heisenberg spaces: The different allowed values of $n_0$ for a given $m$.}\label{table}
\end{table}
For each of these values there is a unique Heisenberg space, except
for $m=3  ~(\mathrm{mod} ~4)$ where there can be many non-equivalent
Heisenberg spaces for a given dimension (see \cite{GHG} for details). 

In particular, the generalised Heisenberg  algebras are 2-step nilpotent; i.e. $[[\mf{n},\mf{n}],\mf{n}]=0$. As the Lie exponential
map, $\exp_{\mf{n}}: \mf{n}\mapsto\mf{H}_{m,n}$, is a diffeomorphism,
the Heisenberg group, $\mf{H}_{m,n}$, will also be a 2-step nilpotent
group. However, note that not all 2-step nilpotent groups are generalised
Heisenberg groups\footnote{For example, the 2-step nilpotent algebra
  $A_{5,1}$ in \cite{PSW} is not a generalised Heisenberg algebra.}.  

\subsubsection*{Damek-Ricci spaces}
Using these generalised Heisenberg groups we can construct other group
manifolds which are called \emph{Damek-Ricci spaces}. These spaces are
 group manifolds similar to the complex hyperbolic spaces and have an Iwasawa decomposition where the
generalised Heisenberg algebras will appear \cite{CDKR}. The Damek-Ricci
space $S_{m,n}$, is defined as the manifold having the (globally
defined) orthonormal
left-invariant one-forms \cite{GHG}
\beq
{\mbold\omega}^A &=&e^{-w}\left(\dx^A+\frac 12B^A_{ab}y^a{\bf dy}^b\right), \qquad
A=1...m \nonumber \\
{\mbold\omega}^a &=&e^{-w/2}\dy^a, \qquad a=1...n. \nonumber \\
{\mbold\omega}^w &=& {\bf dw}.
\eeq 
Hence, they are of dimension $(n+m+1)$. Furthermore, the Damek-Ricci
spaces are Einstein manifolds with negative curvature \cite{Bog}:
\beq
R_{\mu\nu}=-\left(m+\frac n4\right)g_{\mu\nu}.
\eeq
Thus they are spaces obeying (\ref{metricM}). 

As explained earlier, these Damek-Ricci spaces are homogeneous group manifolds of solvable
type. The derived Lie algebra, $[\mf{s},\mf{s}]=\mf{n}$, is of
generalised Heisenberg type.  

Note that the case $m=1$ we have $S_{1,2n}\cong \mb{H}_{\mb{C}}^{n+1}$, so
these spaces generalise the complex hyperbolic spaces. Moreover, the
quaternionic hyperbolic space $\mb{H}_{\mb{H}}^{n+1}$, is one of
the $S_{3,4n}$ spaces, and the Cayley hyperbolic plane,
$\mathsf{Cay}\mb{H}^2$, is isometric to  the $S_{7,8}$
space. Hence, all of the division algebras can be realised in these
spaces\footnote{For more about these spaces and their relatives,
  consult for example \cite{Heber,Damek1,Damek2,Wolter}.}.

\subsection{$\dim(\mf{a})=1$, and $\mf{n}$ nilpotent} 
\label{Nil4}
If we allow for the $\mf{n}$ to be a more general nilpotent algebra,
many more possibilities arise. Unfortunately, not all nilpotent
algebras are known; however, all nilpotent algebras up to dimension 7
are given in \cite{Magnin}. Other examples, such as some infinite
series of nilpotent algebras, are given in \cite{Wolter}. 

For all these spaces, the metric can be written in horospherical
coordinates where the horospheres are \emph{nilgeometries},
${\sf Nil}^n$. As an example, we can consider the space having ${\sf Nil}^4$
horospheres\footnote{There is only one irreducible 4-dimensional real Lie
algebra which is nilpotent. This algebra is three-step nilpotent and
therefore it is  not a generalised Heisenberg algebra.}. Let ${\sf
  Nil}^4$ have the orthonormal left-invariant one-forms 
\beq
{\mbold\omega}^1={\bf dv}, \qquad {\mbold\omega}^2=\dx-y{\bf
  dv},\nonumber \\
{\mbold\omega}^3=\dy-z{\bf dv}, \qquad {\mbold\omega}^4=\dz.
\eeq
Then the metric of the form (\ref{metricM}) with ${\mbold\omega}^i$
given as above, and 
\beq
(p_1,p_2,p_3,p_4)=\left(\frac{1}{2\sqrt{5}},~\frac{2}{\sqrt{5}},~\frac{3}{2\sqrt{5}},~\frac{1}{\sqrt{5}}\right).
\eeq
is an Einstein space with $R_{\mu\nu}=-(3/2)g_{\mu\nu}$. 
\subsection{$\dim(\mf{a})>1$, and $\mf{n}$ nilpotent}
\label{Sol}
In this case, the horospheres are solvmanifolds themselves. All of these cases are not known, but one can find
some particular examples of such spaces. 

For example, there are
solutions as follows. Consider the solvegeometry having the
orthonormal left-invariant one-forms
\beq
&&{\mbold\omega}^n={\bf dv}, \qquad
    {\mbold\omega}^i=e^{-q_iv}\dx^i,\quad 
i=1,...,(n-1), \nonumber \\
&& \sum_{i=1}^{n-1}q_i=0.
\eeq
Then the metric of the form (\ref{metricM}) with ${\mbold\omega}^i$
given as above, and 
\beq
&& p_1=p_2=...=p_{n-1}\equiv p, \quad p_n=0 \\
&& p^2=\frac{1}{n-1}\sum_{i=1}^{n-1}q_i^2 \nonumber
\eeq
is an Einstein space with $R_{\mu\nu}=-(n-1)p^2g_{\mu\nu}$. 

\section{Black Holes} 
As we  now iteratively have constructed spaces for which the horospheres
are products of generalised Heisenberg spaces, nilgeometries, solvegeometries and  a Ricci-flat space, one might
wonder if there are similar generalisations of black
holes. Indeed there are. 

Assume there is a negatively curved Einstein space given by 
\beq
ds^2&=&-e^{-2pw}\dt^2+{\bf dw}^2+\sum_{i=1}^ne^{-2q_iw}\left({\mbold\omega^i}\right)^2,
\qquad R_{\mu\nu}=-\alpha^2g_{\mu\nu},
\label{metricGHBH}\eeq
(here, $p$ must be $p=(\sum q^2_i)/(\sum q_i)$). Then there is a corresponding ``black hole'' spacetime for which the
metric takes the form
\beq
ds^2&=&-e^{-2pw}F(w)\dt^2+\frac{{\bf
  dw}^2}{F(w)}+\sum_{i=1}^ne^{-2q_iw}\left({\mbold\omega^i}\right)^2,
\label{BlackHolemetric}\\ 
F(w)&=& 1-M\exp\left[\left(p+\sum_{i=1}^{n}q_i\right)w\right], \quad 
R_{\mu\nu}=-\alpha^2g_{\mu\nu}.\nonumber 
\eeq
Hence, by doing a proper identification we can from the above
construct black hole spacetimes which have horizon geometries modelled
on \[ \mc{S}\cong \mc{R}^m\times \mathcal{M}_1\times \mathcal{M}_2\times
\cdots \times \mathcal{M}_k, \]
where $\mc{R}^m$ is an $m$-dimensional  Ricci-flat manifold and each $\mathcal{M}_i$ is a generalised Heisenberg group, nilgeometry or a
solvegeometry,  
as follows: we choose a discrete group $\Gamma\subset\isom{(\mc{S})}$ which
acts freely and properly discontinuously on $\mc{S}$, and construct the
quotient $\mc{S}/\Gamma$. Locally, as $w\rightarrow -\infty$, all these black hole spacetimes
asymptotically approach the spacetime given by eq.(\ref{metricGHBH}). 

Here a comment is required. One usually assume that the horizon of
the black hole has finite volume (possibly also compact). Not all
geometries of the above class allow for a finite-volume quotient;
i.e. that $\mc{S}/\Gamma$ has finite volume. Assuming a finite-volume
horizon, we have to restrict ourselves to model geometries in the
sense of Thurston \cite{thurston}:
\paragraph*{Model Geometry:} {\sl A pair $(X,G)$ with $X$ a connected
and simply connected manifold, and $G$ a Lie group acting transitively
on $X$, is called a \emph{model geometry} if the following conditions
are satisfied:
\begin{enumerate}
\item $X$ can be equipped with a $G$-invariant Riemannian metric
\item $G$ is maximal; i.e. there does not exist a larger group $H\supset G$ where  $H$ acts transitively on $X$ and requirement 1 is satisfied. 
\item There exists a discrete subgroup $\Gamma\subset G$ such that
$X/\Gamma$ has finite volume.
\end{enumerate}
}
The model geometries in dimension 3 were found by Thurston
\cite{thur:97,thurston} and are
usually called  ``the 8 Thurston geometries''. The
four-dimensional model geometries were found by Filipkiewicz
\cite{filip} (see also \cite{Wall,Hillman}). Hence, by allowing only model
geometries as the geometry of the horospheres, we ensure that there
exists a $\Gamma\subset\isom(\mathcal{S})$ such that $\mc{S}/\Gamma$
has finite volume.\footnote{The 5-dimensional black hole solutions
  with horizon geometry modelled on the 3-dimensional model geometries
  were considered in \cite{CW}. See also \cite{Kodama1} in this respect.}

Note that the horizons of these black holes are not Einstein manifolds
which  implies that the analysis done in
\cite{GH} is not applicable for these black holes. However, it would be 
interesting to do a similar analysis and check whether or not these black
holes are stable.  
\subsection{6-dimensional Black Holes modelled on solvegeometries.} 
Let us consider a 6-dimensional example, where the horizon is modelled
on the infinite series of model geometries called\footnote{There are 2
  other solvegeometries in dimension 4 which are model geometries;
  namely the ones called $\mathsf{Sol}^4_0$ and $\mathsf{Sol}^4_1$. In
  the notation of Patera \textit{et al.} \cite{PSW,PW}, $\mathsf{Sol}^4_1$ has the
  simply transitive group $A_{4,8}$ while $\mathsf{Sol}^4_0$ has
  $A^{-1/2,-1/2}_{4,5}$ and the infinite series $A^{-2q,q}_{4,6}$ \cite{sigEssay}.}
$\mathsf{Sol}^4_{m,n}$. 

The solvable Lie groups $\mathsf{Sol}^4_{m,n}$ can be considered as $\mathsf{Sol}^4_{m,n}=\mb{R}^3\ltimes_{A}\mb{R}$ where $A$ is the matrix
\beq
A=\exp\begin{bmatrix} 
at & 0 & 0 \\ 0 & bt & 0 \\ 0 &0& ct
\end{bmatrix}.
\eeq
Here, $a>b>c$, $a+b+c=0$, and $\lambda_i=e^a,e^b,e^c$ are the roots of
the cubic
\beq
\lambda^3-m\lambda^2+n\lambda-1=0
\eeq
with $m,n$ positive integers. Note that if $m=n$, we have
$\lambda_2=1$ and thus $\mathsf{Sol}^4_{m,m}=\mathsf{Sol}^3\times
\mb{E}^1$. Proportional matrices $A$ have isomorphic
geometries. An invariant metric is\footnote{This geometry corresponds to the Lie algebra $A_{4,5}^{p,q}$ with
$p=b/a$ and $q=c/a$ in the notation of Patera \textit{et al.} \cite{PSW,PW}.}
\beq
d\sigma^2={\bf dv}^2+e^{av}\dx^2+e^{bv}\dy^2+e^{cv}\dz^2.
\eeq
Hence, this metric is included in the example in section
\ref{Sol}. This means that by using this example, and the Einstein
metric
\beq
\dw^2-e^{-2qw}\dt^2,
\eeq
we can construct a 6-dimensional  Einstein space by 
\beq
M(\mb{E}^1)\odot M(\mathsf{Sol}^4_{m,n})=M(\mb{E}^1\times
\mathsf{Sol}^4_{m,n}). 
\eeq
By using the corresponding metric (\ref{BlackHolemetric}) we can 
construct a black hole metric where the horizon is modelled on the
geometries $\mathsf{Sol}^4_{m,n}$. 

This is only one example of the possible black holes one can
construct in 6 dimensions. Using, for example, the example in section
\ref{Nil4} we can 
similarly construct 6-dimensional black holes modelled on
$\mathsf{Nil}^4$. 

\section{Summary} 
In this paper we have systematically constructed negatively curved
Einstein spaces of various dimensions. The spaces can be classified in
terms of the geometry of the horospheres, and by using a set of building
blocks we gave an iterative procedure of constructing
higher-dimensional Einstein spaces for which the geometry of the
horospheres was an arbitrary product of generalised Heisenberg
spaces, nilgeometries, solvegeometries,
and a Ricci-flat space. The building blocks could be any of the
homogeneous Einstein solvmanifolds. We also showed that all of these
spaces have black hole analogues by explicitly writing down the
metrics. These black holes provide us with an infinite series of
topologically distinct black holes in higher dimensions. The horizon
of these black 
holes are not Einstein in general, however, the non-Einstein part is
always modelled on the so-called model geometries. 

Here in this paper, only a specific type of black holes were
considered. However, the 
Einstein metrics also allow for BTZ black holes \cite{btz,Mann}. For
example, using the metric (\ref{metricHn}) with $m=1$, we can identify
points under the action of $\phi_{\lambda}\circ{\sf A}$, where $
\phi_{\lambda}$ is the dilation and ${\sf A}\in U(n-1)$, to create a
black hole. By performing this identification we create BTZ black
holes similar to those found by Ba\~{n}ados \cite{Banados}. These
BTZ constructions have not been explored in this paper, but 
it would certainly be interesting to investigate these BTZ
analogues as well. 

This paper has set the scene for investigation of higher-dimensional
black holes with horizon geometries which are not Einstein. Several
unanswered questions still remain. For example, are these black holes
stable? What role does the boundary of these spaces play for the
physics in the interior? Is there an AdS/CFT version for these spaces?
So far, none of these questions have been investigated. Only time, and
some more work, can tell what their answers might be. 

\section*{Acknowledgments}
The author would like to thank Paul Davis, Gary W. Gibbons and Hari K. Kunduri for
reading through the manuscript and making useful comments. 
This work was funded by the Research Council of Norway and an Isaac Newton
Studentship.

\appendix
\section{Curvature tensors}
Consider the metric 
\beq
ds^2&=&-e^{-2pw}e^{2\beta}\dt^2+e^{-2\beta}{\bf
dw}^2+\sum_{i=1}^ne^{-2q_iw}\left({\mbold\omega^i}\right)^2, 
\eeq
where $\beta=\beta(w)$. Assume also that the metric with $\beta=0$ is
\emph{homogeneous} where $e^{-q_iw}{\mbold\omega^i}$ are
left-invariant one-forms, and that the space with metric
\beq
\widetilde{ds}^2=\sum_{i=1}^n\left({\mbold\omega^i}\right)^2
\eeq
is homogeneous and  has the Riemann and Ricci tensors given by $\widetilde{R}^i_{~jkl}$
and $\widetilde{R}_{ij}$, respectively. These spaces include most of
the cases considered in this paper. 

In the orthonormal frame, the independent components of the Riemann
tensor are given by (no sum unless explicitly specified)
\beq
R^t_{~wtw}&=&-(\beta''+2(\beta')^2-3p\beta'+p^2)e^{2\beta} \nonumber \\
R^t_{~iti}&=&-q_i(p-\beta')e^{2\beta} \nonumber \\
R^i_{~wiw}&=&-q_i(q_i-\beta')e^{2\beta} \nonumber \\
\underset{i\neq
j}{R^i_{~jij}}&=&\widetilde{R}^i_{~jij}-q_iq_je^{2\beta} \nonumber \\
\text{rest of } {R^i_{~jkl}}&=&\widetilde{R}^i_{~jkl}.
\eeq
The Ricci tensor is thus
\beq
R_{tt}&=&
\left[\beta''+2(\beta')^2-3p\beta'+p^2+(p-\beta')\sum_{i=1}^nq_i\right]e^{2\beta}
\nonumber \\
R_{ww}&=&-
\left[\beta''+2(\beta')^2-3p\beta'+p^2+\sum_{i=1}^nq_i(q_i-\beta')\right]e^{2\beta}
\nonumber \\
R_{ij}&=& \widetilde{R}_{ij}-\delta_{ij}q_i\left(\sum_{k=1}^nq_k+p-2\beta'\right)e^{2\beta}.
\eeq
For an Einstein space we have $R_{tt}+R_{ww}=0$,
which implies $p=(\sum q_i^2)/(\sum q_i)$. Note also that the
non-black hole solutions are given by $\beta'=\beta=0$.

\end{document}